\documentclass[aps,pra,twocolumn,amsmath,amssymb,showpacs,letterpaper]{revtex4}

\usepackage{lmodern}
\usepackage{graphicx}
\usepackage{siunitx}
\usepackage{psfrag}
\usepackage{xspace}
\usepackage{amsmath}
\usepackage{empheq}
\usepackage{mathtools}

\newcommand{\bm}[1]{\boldsymbol{\mathbf{#1}}}
\newcommand{\tens}[1]{\bm{#1}}
\newcommand{\ud}{\mathrm{d}}
\newcommand{\bra}{\left\langle}
\newcommand{\ket}{\right\rangle}

\newcommand{\im}{\operatorname{Im}}
\newcommand{\tr}{\operatorname{Tr}}

\newcommand{\ie}{i.e.\@\xspace}
\newcommand{\eg}{e.g.\@\xspace}

\newcommand{\eq}[1]{Eq.~\eqref{#1}}

\newcommand{\fig}[1]{Fig.~\ref{#1}}

\begin{document}

   \title{Spatial correlations of the spontaneous decay rate as a probe of dense and correlated disordered materials}
   \author{Olivier Leseur}
   \affiliation{ESPCI Paris, PSL Research University, CNRS, Institut Langevin, 1 rue Jussieu, F-75005, Paris, France}
   \author{Romain Pierrat}
   \affiliation{ESPCI Paris, PSL Research University, CNRS, Institut Langevin, 1 rue Jussieu, F-75005, Paris, France}
   \author{R\'emi Carminati}\email{remi.carminati@espci.fr}
   \affiliation{ESPCI Paris, PSL Research University, CNRS, Institut Langevin, 1 rue Jussieu, F-75005, Paris, France}

   \begin{abstract}
      We study theoretically and numerically a new kind of spatial correlation for waves in disordered
      media. We define $C_{\Gamma}$ as the correlation function of the fluorescent decay rate of an
      emitter at two different positions inside the medium. We show that the amplitude and the width of $C_{\Gamma}$
      provide decoupled information on the structural correlation of the disordered medium and on the local environment of the
      emitter. This result may stimulate the emergence of new imaging and sensing modalities in complex media.
   \end{abstract}

   \maketitle

   \section{Introduction}\label{intro}

   Sensing and imaging are key applications of the study of light propagation in strongly scattering
   media~\cite{SEBBAH-2001}. Optical coherence tomography~\cite{HUANG-1991} is one of the most emblematic example but is
   limited to small optical thicknesses where the single scattering regime takes place. Because of multiply scattered
   light, sensing and imaging deeply inside a strongly disordered system is very challenging and has been a matter of
   intense study in the last two decades. Important breakthroughs were achieved recently by ``learning'' the system
   using wavefront shaping techniques~\cite{VELLEKOOP-2007,CARMINATI-2010-4}, by using multimodal approaches such
   acousto-optics~\cite{MARKS-1993}, or by taking advantage of particular features of light scattering in complex
   environments such as the memory effect~\cite{MOSK-2012a}, to cite a few examples.

   Another possibility consists in using fluorescent emitters embedded inside the scattering medium. It is well known
   that the spontaneous decay rate $\Gamma$ of such an emitter strongly depends on the local
   environment~\cite{PURCELL-1946}. More precisely, this decay rate is proportional to the Local Density of States
   $\rho$ (LDOS) at the position of the emitter~\cite{JOULAIN-2003,CARMINATI-2015}. This makes this quantity strongly
   non-universal which is of great interest in terms of imaging~\cite{YODH-1996},
   sensing~\cite{CARMINATI-2010-7,LODAHL-2010,ARXIV-DE_SOUSA-2016} and control~\cite{CHANCE-1978}. By performing
   statistics, signatures of the local order around the emitter~\cite{CAZE-2010} and of transport
   regimes~\cite{PIERRAT-2010} are revealed.

   Interestingly, the fluctuations of the LDOS are encoded in the spatial intensity correlation function (speckle
   correlation) measured outside the medium~.  More precisely, LDOS fluctuations generate an infinite-range contribution
   to the speckle correlation function denoted by $C_0$~\cite{SHAPIRO-1999,TIGGELEN-2006,CAZE-2010}.  This contribution
   is a feature of speckle patterns produced by a point source (\eg a fluorescent emitter) located inside the medium.
   Measuring $C_0$ amounts to measuring LDOS fluctuations. In the optical regime, this can be achieved by measuring
   fluctuations of the spontaneous decay rate of fluorescent emitters~\cite{SAPIENZA-2011,MOSK-2010-1,GARCIA-2012}. In
   acoustics, direct measurements of $C_0$ from speckle correlations have been reported~\cite{HILDEBRAND-2014}.

   In this paper, we introduce and study a new type of spatial correlation function, denoted $C_{\Gamma}$, and defined
   as the correlation function of the spontaneous decay rate of a single emitter measured at two different positions
   inside the disordered medium. As will be shown, this correlation function generalizes the usual $C_0$ contribution.
   We demonstrate that the amplitude and the width of $C_{\Gamma}$ provide decoupled information on the structural
   correlation of the disordered medium and on the local environment of the emitter, which makes this correlation
   function particularly interesting for sensing and imaging in complex media.

   \section{Decay rate statistics and speckle correlations}\label{decay_rate}

   The normalized correlation function of the intensity measured in the far-field is defined as
   \begin{equation}
      C(\bm{u},\bm{u}')=\frac{\bra I(\bm{u})I(\bm{u}')\ket}{\bra I(\bm{u})\ket\bra I(\bm{u}')\ket}-1
   \end{equation}
   where $I(\bm{u})$ is the intensity in direction $\bm{u}$ and $\bra\ldots\ket$ denotes an average over all
   possible configurations of disorder. This correlation function is usually splitted into three
   components~\cite{FENG-1988}
   \begin{equation}
      C(\bm{u},\bm{u}')=C_1(\bm{u},\bm{u}')+C_2(\bm{u},\bm{u}')+C_3(\bm{u},\bm{u}').
   \end{equation}
   The $C_1$ term is usually the predominant short-range term and gives typically the size of the speckle spot. $C_2$
   and $C_3$ are long-range terms with smaller amplitudes. When the speckle pattern is produced by a point source
   embedded inside the scattering medium, an additionnal term of infinite-range exists, and is denoted by
   $C_0$~\cite{SHAPIRO-1999}. The $C_0$ contribution to the correlation function is related to the normalized
   fluctuations of the LDOS $\rho(\bm{r}_0)$~\cite{TIGGELEN-2006,CAZE-2010}
   \begin{equation}
      C_0=\frac{\bra \rho^2(\bm{r}_0)\ket}{\bra\rho(\bm{r}_0)\ket^2}-1
   \end{equation}
   where $\bm{r}_0$ is the position of the emitter. For a fluorescent emitter in the weak-coupling regime, the
   spontaneous decay rate $\Gamma(\bm{r}_0)$ is proportionnal to the LDOS~\cite{CARMINATI-2015}, and $C_0$ can be
   rewritten as
   \begin{equation}\label{C_0}
      C_0=\frac{\bra \Gamma^2(\bm{r}_0)\ket}{\bra\Gamma(\bm{r}_0)\ket^2}-1.
   \end{equation}
   Several studies of LDOS fluctuations in disordered media, or equivalently of $C_0$, have been
   reported~\cite{FROUFE-2007,CARMINATI-2008,PIERRAT-2010,CARMINATI-2010-7,MOSK-2010-1,SAPIENZA-2011,CARMINATI-2015,DOGARIU-2015}.
   It was shown that $C_0$ originates from near-field interactions with the nearby scatterers, providing a non-universal
   behavior that is particularly relevant for sensing and imaging~\cite{CAZE-2010,SAPIENZA-2011}. Moreover, $C_0$ is
   also expected to be influenced by structural correlations in the disorder~\cite{CAZE-2010,ARXIV-DE_SOUSA-2016}.  In
   this article, we show that the new correlation $C_\Gamma$ carries enough information to extract signatures of both
   the near-field interactions and the structural correlations, thus providing a potentially useful extension of the
   usual $C_0$ correlation function.

   \section{Numerical study}\label{numerics}

   \subsection{Methodology}\label{numerics_method}

\begin{figure}[!htb]
      \centering
      \psfrag{L}[c]{$2R$}
      \psfrag{a}[c]{$a$}
      \psfrag{r}[c]{$\bm{r}_0$}
      \psfrag{s}[c]{$\bm{r}_0'$}
      \psfrag{R}[c]{$R$}
      \psfrag{S}[c]{$R_0$}
      \psfrag{d}[c]{$\Delta$}
      \includegraphics[width=0.9\linewidth]{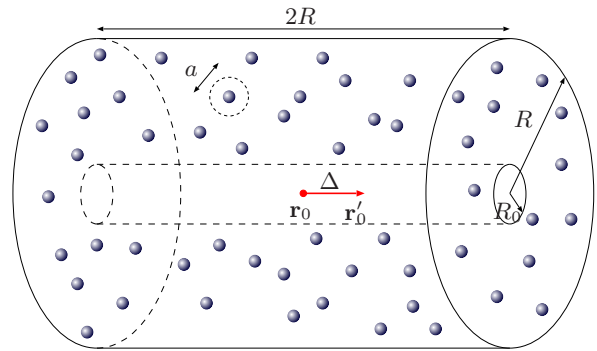}
      \caption{Sketch of the system. The strongly scattering medium lies between two cylinders of radii $R_0$ and $R$,
      respectively, and of length $2R$. The inner region with radius $R_0$ corresponds to the region within the medium
      in which the fluorescent source is free to move, from position $\bm{r}_0$ to position  $\bm{r}_0'$.  To mimic hard
      spheres correlations, a minimum distance $a$ is forced between scatterers.}
      \label{cylinder}
   \end{figure}

   To get insight into the behavior of the new correlation function, we begin with a numerical study. The system of
   interest is depicted in \fig{cylinder}. $N$ point-dipole scatterers are lying between two coaxial cylinders of radii
   $R_0$ and $R$ respectively and of longitudinal size $2R$. The inner region with radius $R_0$ corresponds to the
   region within the medium in which the fluorescent source is free to move. The optical properties of the scatterers
   are described by an electric polarizability
   \begin{equation}\label{alpha}
      \alpha(\omega)=-\frac{6\pi\gamma c^3}{\omega_0^2}\frac{1}{\omega^2-\omega_0^2+i\gamma\omega^3/\omega_0^2}
   \end{equation}
   where $\omega$ is the emission frequency, $\omega_0$ the resonant frequency of the scatterers, $\gamma$ the linewidth and $c$ the speed of light
   in vacuum. From the polarizability $\alpha$, we can compute the scattering ($\sigma_s$) and the extinction
   ($\sigma_e$) cross-sections of one scatterer. They are given by
   \begin{equation}\label{sigma_s}
      \sigma_s=\frac{k_0^4}{6\pi}\left|\alpha(\omega)\right|^2
      \quad;\quad
      \sigma_e=k_0\im\alpha(\omega)
   \end{equation}
   where $k_0=\omega/c=2\pi/\lambda$. The optical theorem is correctly fulfilled by the polarizability model, and in a non-absorbing medium
   such as the one considered here, we have $\sigma_e=\sigma_s$.  Defining the density of scatterers by
   $\mathcal{N}=N/V$ where $V=2\pi R\left(R^2-R_0^2\right)$ is the volume of the scattering system, we have also access
   to the scattering mean-free path $\ell_B$ in the limit of an uncorrelated system (Boltzmann mean-free path). Its
   expression is
   \begin{equation}\label{l_B}
      \ell_B=\frac{1}{\mathcal{N}\sigma_s}.
   \end{equation}
   To generate disorder correlations, a fictitious exclusion volume of diameter $a$ is forced between scatterers.  This
   mimic a hard sphere potential. By increasing the value of $a$, one increases the correlation level.  Instead of using
   the parameter $a$ to characterize the level of structural correlation, we use the effective volume fraction $f$
   defined as
   \begin{equation}
      f=\frac{NV_0}{V}=\mathcal{N}V_0
   \end{equation}
   where $V_0=\pi a^3/6$ is the exclusion volume around each scatterer. Note that the effective volume fraction $f$ has
   to be understood as a correlation parameter, that is changed by changing $V_0$ only (the real density of scatterers
   $\mathcal{N}$ is constant throughout the study).

   The emitter lies initially at $\bm{r}_0$, the center position (see \fig{cylinder}), that can be changed to another
   position $\bm{r}_0'$ along the cylinder axis. The distance $R_0$ corresponds to the minimum distance forced between
   the source and all scatterers. In other words, it parameterizes the near-field environment of the source (proximity
   of scatterers). The spatial correlation function $C_{\Gamma}$ studied in this paper is defined as
   \begin{equation}\label{c_gamma_expr}
      C_{\Gamma}(\Delta)=\frac{\bra\Gamma(\bm{r}_0)\Gamma(\bm{r}_0')\ket}{\bra\Gamma(\bm{r}_0)\ket\bra\Gamma(\bm{r}_0')\ket}-1
   \end{equation}
   where $\Delta=|\bm{r}_0-\bm{r}_0'|$. For $\Delta=0$, this expression coincides with the definition of $C_0$ in \eq{C_0}.

   To compute $C_{\Gamma}$, we have first to solve Maxwell's equations for a point dipole source. For that purpose, we
   use the coupled dipoles method~\cite{LAX-1952}. It consists in calculating first the exciting field on each scatterer
   given by a set of $N$ linear equations:
   \begin{equation}
      \bm{E}_j=\mu_0\omega^2\tens{G}_0(\bm{r}_j,\bm{r}_0,\omega)\bm{p}
         +\alpha(\omega)k_0^2\sum_{\substack{k=1\\k\ne j}}^N\tens{G}_0(\bm{r}_j,\bm{r}_k,\omega)\bm{E}_k
   \end{equation}
   where $\mu_0$ is the vacuum permeability and $\bm{p}$ the source dipole. $\tens{G}_0$ is
   the Green function in vacuum. For vector waves in three dimensions, it is given by
   \begin{equation}\label{vacuum_green}
      \tens{G}_0(\bm{r},\bm{r}_0)=
         \operatorname{PV}\left\{
            \left[\tens{I}+\frac{\bm{\nabla}_{\bm{r}}\otimes\bm{\nabla}_{\bm{r}}}{k_0^2}\right]\frac{\exp\left[ik_0R\right]}{4\pi
            R}
         \right\}-\frac{\delta\left(\bm{R}\right)}{3k_0^2}\tens{I}
   \end{equation}
   where $\operatorname{PV}$, $\tens{I}$, $\otimes$, $\delta$ are the Cauchy principal value operator, the identity
   tensor, the tensor product operator and the Dirac delta function respectively. We have used the notations $\bm{R}=\bm{r}-\bm{r}_0$ and
   $R=|\bm{R}|$. Once the exciting fields on each scatterer are known, the field at any position can be computed using
   \begin{equation}
      \bm{E}(\bm{r})=\mu_0\omega^2\tens{G}_0(\bm{r},\bm{r}_0,\omega)\bm{p}
         +\alpha(\omega)k_0^2\sum_{k=1}^N\tens{G}_0(\bm{r},\bm{r}_k,\omega)\bm{E}_k.
   \end{equation}
   By varying the orientation of the source dipole $\bm{p}$, the Green function of the full system $\tens{G}$ can be
   obtained from the relation $\bm{E}(\bm{r})=\mu_0\omega^2\tens{G}(\bm{r},\bm{r}_0,\omega)\bm{p}$. 
   The LDOS averaged over all orientations of the source dipole can be deduced by~\cite{CARMINATI-2015}
   \begin{equation}
      \rho(\bm{r}_0,\omega)=\frac{2\omega}{\pi c^2}\im\left[\tr\tens{G}(\bm{r}_0,\bm{r}_0,\omega)\right]
   \end{equation}
   where $\operatorname{Tr}$ denotes the trace of a tensor. We usually prefer to deal with normalized quantities.
   Defining the vacuum LDOS as $\rho_0=\omega^2/\left(\pi^2c^3\right)$ and the vacuum decay rate of the emitter by
   $\Gamma_0$, we have
   \begin{equation}
     \frac{\Gamma(\bm{r}_0)}{\Gamma_0} = \frac{\rho(\bm{r}_0,\omega)}{\rho_0}
         =\frac{2\pi}{k_0}\im\left[\tr\tens{G}(\bm{r}_0,\bm{r}_0,\omega)\right].
   \end{equation}
   Repeating the operation for another position $\bm{r}_0'$ of the source and averaging over disorder configurations
   leads to an estimate of $C_{\Gamma}$.

   \subsection{Numercial results}\label{numerics_results}

\begin{figure*}[!htb]
      \centering
      \psfrag{a}[c]{(a)}
      \psfrag{b}[c]{(b)}
      \psfrag{x}[c]{$\Delta/R_0$}
      \psfrag{C}[b]{$C_{\Gamma}(\Delta)$}
      \psfrag{f1}[l][c]{$f=\num{1e-3}$}
      \psfrag{f2}[l][c]{$f=\num{0.0181}$}
      \psfrag{f3}[l][c]{$f=\num{0.0317}$}
      \psfrag{f4}[l][c]{$f=\num{0.1}$}
      \psfrag{f}[c]{$f$}
      \psfrag{C0}[b]{$C_0$}
      \includegraphics[width=0.7\linewidth]{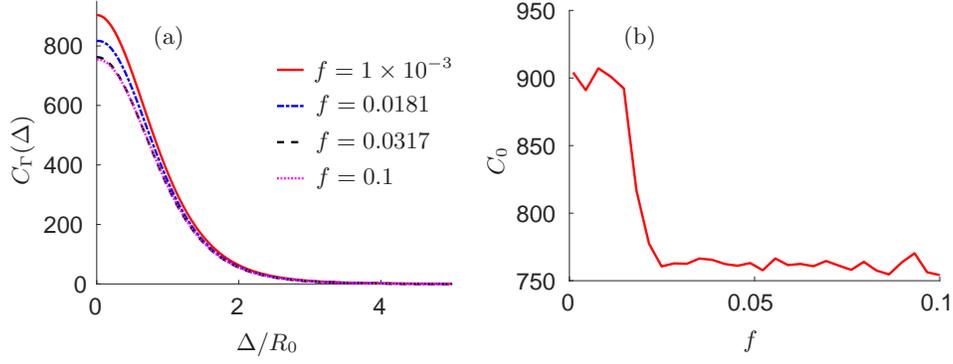}
      \caption{(a) Correlation function $C_{\Gamma}$ of the decay rate as a function of the normalized distance $\Delta/R_0$ for
      four different levels $f$ of structural correlation of the disorder. (b) Amplitude of the spatial decay rate correlation at
      $\Delta/R_0=0$ ($C_0$ correlation) as a function of the disorder correlation level $f$. The parameters are
      $k_0R_0=\num{0.2}$, $k_0R=\num{11.9}$ and $N=100$ with $k_0\ell_B=19$ and $b_B=\num{1.25}$. Depending on the value
      of $f$, between ten and one hundred million configurations are needed to perform the statistical average.}
      \label{c_gamma}
   \end{figure*}

   We have performed numerical simulations on a system with parameters such that $k_0\ell_B=19$ (strength of the
   disorder) and $b_B=2R/\ell_B=\num{1.25}$ (optical thickness). The parameters are given in the caption of
   \fig{c_gamma}.  Thus the numerical simulations are performed in a dilute system and close to the single-scattering
   regime. The simulations are performed by varying the level of structural correlation of the disorder, measured by the
   correlation parameter $f$. The results are shown in \fig{c_gamma}\,(a). We clearly see that the width of the curves
   is independent on $f$, suggesting that the width depends essentially on the microscopic length scale $R_0$ that
   measures the proximity of scatterers around the emitter. The dependence of $f$ is encoded in the amplitude of the
   correlation function $C_{\Gamma}$ for $\Delta=0$, as shown in \fig{c_gamma}\,(b). Note that this amplitude
   corresponds to $C_0$, that is known to depend on $f$~\cite{CAZE-2010}. 
 
   For weak structural correlations of the disorder (small values of $f$), the exclusion distance $a$ between scatterers
   is small, and more than one scatterer can lie in the near field of the emitter. This implies that the emitter can
   interact with two or more scatterers, inducing a strong dependence of the amplitude of $C_\Gamma$ on the correlation
   parameter $f$.  For a high level of structural correlations (large values of $f$), the exclusion distance $a$ is
   large enough to exclude the possibility of interaction with more than one scatterer. For that reason, the amplitude
   of $C_\Gamma$ is almost independent on $f$ in this regime. This qualitatively explains the shape of the curve in
   \fig{c_gamma}\,(b).

   \section{Analytical theory}\label{analytics}

   To get physical insight, we support the numerical data by a theoretical analysis. This has also the advantage to
   provide simple analytical formulas that could be useful in practice.  As the optical thickness $b_B$ is close to
   unity, the system operates in the single scattering regime. In that case, the correlation function can be computed
   analytically, at the price of a few crude but nevertheless controlled approximations.  In the single-scattering
   regime, the decay rate is given by
   \begin{equation}
      \frac{\Gamma(\bm{r}_0)}{\Gamma_0}=1+2\pi
      k_0\sum_{j=1}^N\im\left[\alpha(\omega)\tr\tens{G}_0(\bm{r}_j,\bm{r}_0)^2\right].
   \end{equation}
   Using the expression of the vacuum Green tensor [\eq{vacuum_green}], we find
   \begin{multline}
      \frac{\Gamma(\bm{r}_0)}{\Gamma_0}=1+2\pi k_0\sum_{j=1}^N\im\left[\alpha(\omega)
         \left\{
            2-\frac{10}{k_0^2R_j^2}+\frac{6}{k_0^4R_j^4}
         \right.\right.
   \\\left.\left.
            +i\left(\frac{4}{k_0R_j}-\frac{12}{k_0^3R_j^3}\right)
         \right\}
         \frac{\exp(2ik_0R_j)}{16\pi^2R_j^2}\right]
   \end{multline}
   where $R_j=|\bm{r}_j-\bm{r}_0|$. As $\Delta=|\bm{r}_0-\bm{r}_0'|$ is on the order of $R_0\ll\lambda$, we consider
   that the most important contribution is given by the scatterers lying in the near field of the emitter. Thus we now
   consider a subset $\Omega$ of scatterers located in the vicinity of the source inside a volume $V'$. This subset is
   defined by
   \begin{equation}
      \Omega=\{j\mid \bm{r}_j\in V'\}
   \end{equation}
   and under this near-field approximation, the decay rate becomes
   \begin{equation}
      \frac{\Gamma(\bm{r}_0)}{\Gamma_0}=1+\frac{3\alpha''(\omega)}{4\pi k_0^3}\sum_{j\in\Omega}\frac{1}{R_j^6}
   \end{equation}
   where $\alpha''(\omega)=\im\alpha(\omega)$. The computation of the correlation $C_{\Gamma}$ [\eq{c_gamma_expr}]
   requires the computation of the first two statistical moments of the decay rate:
   \begin{equation}
      \left\{\begin{aligned}
         \bra\Gamma(\bm{r}_0)\ket & =\int
            \Gamma(\bm{r}_0)P(\{\bm{r}_j\})\{\ud^3\bm{r}_j\},
      \\
         \bra\Gamma(\bm{r}_0)\Gamma(\bm{r}_0')\ket & =\int
            \Gamma(\bm{r}_0)\Gamma(\bm{r}_0')P(\{\bm{r}_j\})\{\ud^3\bm{r}_j\}
      \end{aligned}\right.
   \end{equation}
   where $P(\{\bm{r}_j\})$ is the probability density of having the scatterers at positions $\{\bm{r}_j\}$. We denote by
   $N'$ the average number of scatterers in $\Omega$, a quantity that depends essentially on the exclusion radius $R_0$ around
   the emitter and on the correlation parameter $f$. Using this notation, the average decay rate is given by
   \begin{equation}
      \frac{\bra\Gamma(\bm{r}_0)\ket}{\Gamma_0}=1+\frac{3N'\alpha''(\omega)}{4\pi k_0^3}\int_{V'}\frac{P(\bm{r}_j)}{R_j^6}\ud^3\bm{r}_j
   \end{equation}
   where $P(\bm{r}_j)$ is the probability density of finding one scatterer at position $\bm{r}_j$. The integral involves
   a fast decaying function in space, meaning that the integration volume $V'$ can be replaced by $V$ without changing
   the result. Then $P(\bm{r}_j)=V^{-1}$ and we obtain
   \begin{equation}
      \frac{\bra\Gamma(\bm{r}_0)\ket}{\Gamma_0}=1+\frac{3N'\alpha''(\omega)}{4\pi
      k_0^3V}\int\frac{\ud^3\bm{r}}{|\bm{r}-\bm{r}_0|^6}.
   \end{equation}
   The second moment can be obtained in a similar way, leading to
   \begin{multline}
      \frac{\bra\Gamma(\bm{r}_0)\Gamma(\bm{r}_0')\ket}{\Gamma_0^2}
         =\frac{\bra\Gamma(\bm{r}_0)\ket}{\Gamma_0}+\frac{\bra\Gamma(\bm{r}_0')\ket}{\Gamma_0}-1
            +\left[\frac{3\alpha''(\omega)}{4\pi k_0^3}\right]^2
   \\\times
            \left[
               N'\int\frac{P(\bm{r}_j)}{R_j^6{R_j'}^6}\ud^3\bm{r}_j
               +N'(N'-1)\int\frac{P(\bm{r}_j,\bm{r}_k)}{R_j^6{R_k'}^6}\ud^3\bm{r}_j\ud^3\bm{r}_k
            \right]
   \end{multline}
   where $R_k'=|\bm{r}_k-\bm{r}_0'|$ and $P(\bm{r}_j,\bm{r}_k)$ is the probability density of having two scatterers at
   positions $\bm{r}_j$ and $\bm{r}_k$. It is given by
   \begin{equation}
      P(\bm{r}_j,\bm{r}_k)=P(\bm{r}_j)P(\bm{r}_k)\left[1+h(\bm{r}_j,\bm{r}_k)\right]
   \end{equation}
   with $h$ the pair correlation function. This leads to
   \begin{multline}
      \frac{\bra\Gamma(\bm{r}_0)\Gamma(\bm{r}_0')\ket}{\Gamma_0^2}
         =\frac{\bra\Gamma(\bm{r}_0)\ket}{\Gamma_0}\frac{\bra\Gamma(\bm{r}_0')\ket}{\Gamma_0}
            +\left[\frac{3\alpha''(\omega)}{4\pi k_0^3}\right]^2
   \\\times
            \left[
               \frac{N'}{V}\int\frac{\ud^3\bm{r}}{|\bm{r}-\bm{r}_0|^6|\bm{r}-\bm{r}_0'|^6}
               -\frac{N'}{V^2}\int\frac{\ud^3\bm{r}}{|\bm{r}-\bm{r}_0|^6}\int\frac{\ud^3\bm{r}}{|\bm{r}-\bm{r}_0'|^6}
            \right.
   \\
            \left.
               +\frac{N'(N'-1)}{V^2}\int\frac{h(\bm{r},\bm{r}')\ud^3\bm{r}\ud^3\bm{r}'}{|\bm{r}-\bm{r}_0|^6|\bm{r}'-\bm{r}_0'|^6}
            \right]
   \end{multline}
   from which the following expression of the correlation function $C_\Gamma$ is readily deduced:
   \begin{multline}\label{c_gamma_inter}
      C_{\Gamma}(\Delta)
         =\frac{9\alpha''(\omega)^2}{16\pi^2k_0^6}
         \frac{\Gamma_0^2}{\bra \Gamma(\bm{r}_0)\ket\bra \Gamma(\bm{r}_0')\ket}
         \frac{N'}{V}
   \\\times
         \left[
            \int\frac{\ud^3\bm{r}}{|\bm{r}-\bm{r}_0|^6|\bm{r}-\bm{r}_0'|^6}
            -\frac{1}{V}\int\frac{\ud^3\bm{r}}{|\bm{r}-\bm{r}_0|^6}\int\frac{\ud^3\bm{r}}{|\bm{r}-\bm{r}_0'|^6}
         \right.
   \\
         \left.
            +\frac{N'-1}{V}\int\frac{h(\bm{r},\bm{r}')\ud^3\bm{r}\ud^3\bm{r}'}{|\bm{r}-\bm{r}_0|^6|\bm{r}'-\bm{r}_0'|^6}
         \right].
   \end{multline}
   To compute the integrals analytically, we consider infinite cylinders (\ie $R\gg\Delta$) and a small radius for the
   inner cylinder (\ie $R_0\ll R$).  We obtain
   \begin{equation}
      \left\{\begin{aligned}
         \int\frac{\ud^3\bm{r}}{|\bm{r}-\bm{r}_0|^6} & =\frac{\pi^2}{4R_0^3},
      \\
         \int\frac{\ud^3\bm{r}}{|\bm{r}-\bm{r}_0|^6|\bm{r}-\bm{r}_0'|^6} &
            =\frac{\pi^2}{2R_0^3}\frac{\Delta^2+28R_0^2}{\left(\Delta^2+4R_0^2\right)^4}
            \le\frac{7\pi^2}{128R_0^9}.
      \end{aligned}\right.
   \end{equation}
   As $V\gg R_0^3$, the second term in \eq{c_gamma_inter} can be neglected and the correlation $C_{\Gamma}$ reduces to
   the first and the last terms. The bulk pair correlation function is considered such that it only depends on the
   distance between the two points $\bm{r}$ and $\bm{r}'$. We also consider a small correlation level ($f\simeq 0.1$)
   such that the pair correlation function can be approximated by (for higher structural correlation levels, a refined
   model is required~\cite{PERCUS-1958})
   \begin{equation}
      h(|\bm{r}-\bm{r}'|)=\begin{cases}
         -1 & \text{if $|\bm{r}-\bm{r}'|<a$}
      \\
         0 & \text{elsewhere.}
      \end{cases}
   \end{equation}
   This leads to
   \begin{equation}\label{correl_term}
      \int\frac{h(\bm{r},\bm{r}')\ud^3\bm{r}\ud^3\bm{r}'}{|\bm{r}-\bm{r}_0|^6|\bm{r}'-\bm{r}_0'|^6}
         =\int_V\frac{\ud^3\bm{r}}{|\bm{r}-\bm{r}_0|^6}\int_{V_0(\bm{r})}\frac{\ud^3\bm{r}'}{|\bm{r}'-\bm{r}_0'|^6}
   \end{equation}
   where $V_0(\bm{r})$ is the exclusion volume around the scatterer centered at position $\bm{r}$. For dilute media,
   the quantity $1/|\bm{r}'-\bm{r}_0'|^6$ is slowly varying and can be replaced by $1/|\bm{r}-\bm{r}_0'|^6$, so that
   \eq{correl_term} reduces to
   \begin{equation}
      \int\frac{h(\bm{r},\bm{r}')\ud^3\bm{r}\ud^3\bm{r}'}{|\bm{r}-\bm{r}_0|^6|\bm{r}'-\bm{r}_0'|^6}
         \sim V_0\int\frac{\ud^3\bm{r}}{|\bm{r}-\bm{r}_0|^6|\bm{r}-\bm{r}_0'|^6}.
   \end{equation}
   Finally, the correlation function $C_{\Gamma}$ is given by
   \begin{multline}\label{c_gamma_final}
      C_{\Gamma}(\Delta)
         =\frac{9\alpha''(\omega)^2}{32k_0^6R_0^3}\frac{\Gamma_0^2}{\bra \Gamma(\bm{r}_0)\ket\bra \Gamma(\bm{r}_0')\ket}\frac{N'}{V}
         \left[
            1+(N'-1)\frac{V_0}{V}
         \right]
   \\\times
            \frac{\Delta^2+28R_0^2}{\left(\Delta^2+4R_0^2\right)^4}
   \end{multline}
   with
   \begin{equation}
      \frac{\bra\Gamma(\bm{r}_0)\ket}{\Gamma_0}=1+\frac{3\pi\alpha''(\omega)}{16\pi k_0^3 R_0^3}\frac{N'}{V}.
   \end{equation}
   This expression provides a theoretical basis to the qualitative discussion presented at the end of
   section~\ref{numerics}.  First, it shows that the width of the correlation function $C_\Gamma$ depends only on the
   exclusion radius $R_0$, \ie on the minimum distance between the fluorescent source and the nearest scatterers. More
   precisely, the full width at half maxium can be approximated by $\ell_{1/2}=0.89R_0$.  Second, as far as the
   amplitude is concerned, two important regimes can be identified. For large values of $f$ (\ie $f>0.02$), the source
   is chiefly interacting with one scatterer, so that $N'=1$. From \eq{c_gamma_final}, one sees that the amplitude of
   $C_\Gamma$ does not depend on $f$ in this regime, as already seen in \fig{c_gamma}\,(b). Conversely, for small values
   of $f$, the source can interact with more than one scatterer, and the amplitude  of $C_\Gamma$ depends on $f$ through
   both $N'$ and $V_0$.

   \begin{figure*}[!htb]
      \centering
      \psfrag{a}[c]{(a)}
      \psfrag{b}[c]{(b)}
      \psfrag{x}[c]{$\Delta/R_0$}
      \psfrag{C}[b]{$C_{\Gamma}(\Delta)/C_0$}
      \psfrag{N}[l][c]{Numerics}
      \psfrag{A}[l][c]{Analytics}
      \includegraphics[width=0.7\linewidth]{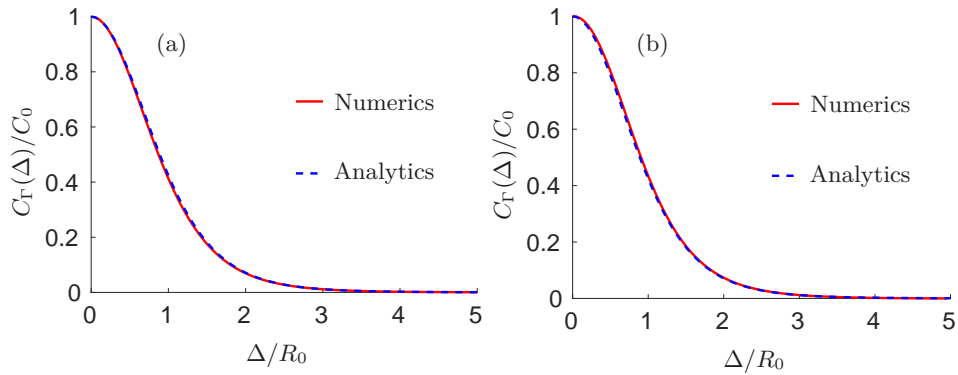}
      \caption{Numerical (red solid line) and analytical (blue dotted line) spatial decay rate correlation $C_{\Gamma}$
      normalized by $C_0$ as a function of the normalized distance $\Delta/R_0$ for $f=\num{1e-3}$ (a) and $f=\num{0.1}$
      (b).  Same parameters as in \fig{c_gamma}.}
      \label{c_gamma_ana}
   \end{figure*}

   It is interesting to compare precisely the numerical results with the approximate analytical model. The comparison is
   shown in \fig{c_gamma_ana}.  It can be seen that \eq{c_gamma_final} describes very well the dependance of the
   correlation function $C_{\Gamma}$ on $\Delta$ for all values of $f$, showing that the analytical model provides a
   very accurate description of $C_{\Gamma}$. 
   
   \section{Conclusion}\label{conclu}

   In summary, we have highlighted a new type of spatial correlation function based on the decay rate of a fluorescent
   emitter measured at two different positions inside a disordered medium. A numerical and analytical study has revealed
   that this correlation function contains more information than the usual $C_0$ intensity correlation function. In
   particular, by measuring its width and amplitude, it is possible to decouple the effect of the near-field
   interactions (proximity effects) and of the structural correlation of disorder. This opens new perspectives for
   imaging and sensing in complex media, and in particular in correlated media whose interest in photonics is growing
   up.

   \begin{acknowledgements}
      This work was supported by LABEX WIFI (Laboratory of Excellence within the French Program ``Investments for the
      Future'') under references ANR-10-LABX-24 and ANR-10-IDEX-0001-02 PSL*.
   \end{acknowledgements}


\end{document}